\documentclass[reprint, amsmath,amssymb, prl, 
]{revtex4-1}
\usepackage{stmaryrd}
\usepackage{amssymb}
\usepackage{}
\usepackage{graphicx}
\usepackage{dcolumn}
\usepackage{bm}
\usepackage{times}
\usepackage{color}

\begin{document}

\title{Bloch oscillations of topological edge modes}
\author{Chunyan Li,${}^{1,2}$ Weifeng Zhang,${}^{2}$ Yaroslav V. Kartashov,${}^{3,4}$ Dmitry V. Skryabin,${}^{5}$ Fangwei Ye${}^{2,6\ast}$\\
\emph{${}^{1}$School of Physics and Optoelectronic Engineering, Xidian University, Xi'an 710071, China}\\
\emph{${}^{2}$Key Laboratory for Laser Plasma (Ministry of Education), Collaborative Innovation Center of IFSA, School of Physics and Astronomy, Shanghai Jiao Tong University, Shanghai 200240, China}\\
\emph{${}^{3}$ICFO-Institut de Ciencies Fotoniques, The Barcelona Institute of Science and Technology,}\\ 	
\emph{ 08860 Castelldefels (Barcelona), Spain}\\
\emph{${}^{4}$Institute of Spectroscopy, Russian Academy of Sciences, Troitsk, Moscow, 108840, Russia}\\
\emph{${}^{5}$Department of Physics, University of Bath, BA2 7AY Bath, UK}\\
\emph{${}^{6}$Department of Physics, Zhejiang Normal University, Jinhua 321004, China}\\
\emph{$^\ast$E-mail:  fangweiye@sjtu.edu.cn}
}
\begin{abstract}
Under the action of a weak constant force a wavepacket in periodic potential undergoes periodic oscillations in space, returning to the initial position after one oscillation cycle. This wave phenomenon, known as Bloch oscillations (BOs), pertains to many physical systems. Can BOs also occur in topological insulators with topologically protected edge states? This question is highly nontrivial, because in topological insulators with broken time-reversal symmetry, the edge states propagate unidirectionally without backscattering, hence BOs that typically involve stages, where wavepacket moves along and against the direction of the force, seem to be impossible in such systems when force acts parallel to the edge of the insulator. Here we reveal that BOs still occur with topological edge states, but in a nonconventional way: they are accompanied not only by oscillations along the edge in the direction of force, but also by oscillations in the direction transverse to that of the force. Full BO cycle involves switching between edge states at the opposite edges through delocalized bulk modes. Bloch oscillations of the topological edge states require to scan the first Brillouin zone twice to complete one cycle, thus they have a period that is two times larger than the period of usual BOs. All these unusual properties are in contrast to BOs in non-topological systems.
\end{abstract}

\maketitle

\section{\uppercase\expandafter{\romannumeral1}.Introduction}
Bloch oscillations (BOs) were introduced in seminal works that addressed the electron dynamics in crystalline lattices under the action of a constant electric field~\cite{bloch, zener}. They were observed for electrons in semiconductor superlattices~\cite{semicon1, semicon2}, shortly after the observation of Wannier-Stark ladders~\cite{ladder1,ladder2}. As a universal wave phenomenon, BOs have been shown to occur in a variety of physical systems, including ultracold atoms~\cite{atom1, atom2}, Bose-Einstein condensates trapped in optical lattices~\cite{BEC1, BEC2}, waveguide arrays~\cite{wg1,wg2,wg3,frac1,frac2}, optically induced lattices~\cite{lattice1,lattice2}, surface plasmon waves in plasmonic waveguides~\cite{spp}, and parity-time symmetric systems~\cite{pt}.

In contrast to conventional insulators, topological insulators~\cite{review1, review2} conduct at the edges of the structure and insulate in the bulk. The edge conductance is due to the existence of in-gap states that are spatially localized at the boundaries, propagate unidirectionally, and are immune to scattering by perturbations or disorder. Such a robustness is a consequence of topological protection~\cite{review1, review2}. The first observations of topological insulator states were performed in electronic systems and recently studies have been extended to electromagnetic waves~\cite{photonics}. Topological edge states have been proposed and observed in gyromagnetic photonic crystals~\cite{gyro1,gyro2}, semiconductor quantum wells~\cite{Lindner2011Floquet}, arrays of coupled resonators~\cite{resonator1, resonator2}, metamaterial superlattices~\cite{supperlattice}, helical waveguide arrays~\cite{helical1,helical2}, and in polariton microcavities, where strong photon-exciton coupling leads to the formation of half-light half-matter polariton quasi-particles~\cite{pol1,pol2,pol3,pol4,pol4a,pol5,pol6}, see \cite{pol6a} for recent experimental realization of polariton topological insulator. Remarkably, application of a constant force to unbounded topological systems has been shown to provide a powerful tool for the measurement of its topological invariants~\cite{unbound1,unbound2,unbound3,unbound4}. It should be stressed that none of these works addressed Bloch oscillations in truncated topological insulators.

Here we combine the physics of topological insulators with BOs and introduce unusual BOs of topological edge states. Taking into account their unidirectional propagation nature at the edges, topological edge states at the first glance cannot undergo BOs, because the latter implies that the wavepacket periodically returns to its initial position. Our study reveals that the BOs of topological edge states are still possible, but in form that sharply contrasts with BOs in the non-topological systems: although they are not able to propagate back and forth along the same side to complete BOs, topologial edge states still manage to restore their initial position periodically by switching into their counterpart at the other side of the structure that is propagating in the opposite direction. Thus, the BOs of topological edge state involves not only longitudinal oscillations along the gradient, but also involves oscillations between the two edges of the topological insulator - which do not occur in non-topological systems. Further, in corresponding momentum space the wavepacket evolution in the course of BOs connects the energy bands surrounding the topological gap and therefore induce nearlly complete interband transitions even for very small gradients. This is in contrast to the usual Landau-Zener tunneling that occurs in non-topological systems, which remains non-complete even for large gradients~\cite{bo1}.

In this work we address BOs emerging in \textit{truncated} topological insulators based on the microcavity exciton-polaritons model, however, the results can be carried over to a variety of other topological systems. BOs in topological systems may be investigated also in helical waveguide arrays with topological gap~\cite{blhelix}, especially if radiative losses can be reduced to an extent that a complete oscillation cycle can be observed. Also, some mathematical aspects of topological Bloch oscillations on infinite lattices, not involving edge states, are discussed in \cite{mathtop}.

\section{\uppercase\expandafter{\romannumeral2}.Model}

 To provide an example of topological system, where BOs are possible, we consider exciton-polariton topological insulator based on microcavity whose top mirror is structured into a honeycomb lattice truncated on two sides. The resulting structure is a honeycomb lattice ribbon with zigzag edges in one direction and infinite along the other direction. The linear potential inducing Bloch oscillations features a gradient parallel to the ribbon edges. For the experimental realization of such a gradient in a polaritonic microcavity see, e.g., Ref.~\cite{pol7}.

We address the evolution of a spinor polariton wavefunction $\boldsymbol{\psi}=(\psi_{+},\psi_{-})^\textrm{T}$ governed by the coupled Schr\"{o}dinger equations \cite{pol1,pol6}:
\begin{equation}
\begin{split}
i \frac{\partial \boldsymbol{\psi}}{\partial t}& =-\frac{1}{2}(\partial_{x}^2+ \partial_{y}^2) \boldsymbol{\psi}
+\sigma_{1} \beta (\partial_{x} \mp i\partial_{y})^2 \boldsymbol{\psi}\\
& +\sigma_{3}\Omega \boldsymbol{\psi}+[\mathcal{R}(x,y)+\alpha y]\boldsymbol{\psi},
\end{split}
\label{eq:SEs}
\end{equation}
where $x,y$ are the coordinates scaled to the characteristic length $x_{0}$; all energy parameters (such as the potential depth and the Zeeman splitting) are scaled to $\varepsilon _{0}=\hbar^{2}/mx_{0}^{2}$, where $m$ is the effective polariton mass that corresponds to the gradient and periodic potential free system; $t$ is the evolution time scaled to $t_{0}=\hbar\varepsilon _{0}^{-1}$; $\psi_{+}$ and $\psi_{-}$ are the spin-positive and spin-negative components of the wavefunction in the circular polarization basis \cite{pol1}; $\sigma_1, \sigma_3$ are the Pauli matrices; $\beta$ is the strength of spin-orbit coupling arising from the fact that tunneling between microcavity pillars is polarization-dependent; $\Omega$ is the Zeeman splitting; the potential landscape $\mathcal{R}(x,y)=-p\sum_{m,n}\mathcal{Q}(x-x_m,y-y_n)$ created by microcavity pillars arranged into a honeycomb array with nodes at the points $(x_m, y_n)$ is composed of Gaussian wells $\mathcal{Q}(x,y)=\textrm{exp}[-(x^2+y^2)/d^2]$ of characteristic diameter $2d$, depth $p$, and separation $a$ between neighboring wells; the parameter $\alpha$ describes a small gradient along $y$ that is necessary for occurrence of BOs. We assume that the array of microcavity pillars is periodic along the $y$-axis with a period $Y=3^{1/2}a$ and that it is truncated along the $x$-axis in such a way that the topological insulator acquires two zigzag edges [see two $y$-periods of this structure in Fig.~1(a)]. A potential gradient is applied along the edges of the insulator. For $x_{0}=2~\mu \text{m}$ and effective mass $m=10^{-34} ~\text{kg}$, one gets $\varepsilon _{0}\approx 0.17~ \text{meV}$ and $t_{0}\approx 3.8~ \text{ps}$. We set $p=8$, which corresponds to $1.38 ~\text{meV}$, $d=0.4$ ($1.6~\mu \text{m}$ diameter), and $a=1.4$ ($2.8 ~\mu \text{m}$ separation between pillars). A potential energy gradient in the microcavity can be created by slight variations of its thickness along the $y$-axis, as realized experimentally in \cite{pol7}. Here we consider gradients in the range $\alpha=0.001-0.01$ (0.04 meV/mm-0.4 meV/mm). Note that the potential gradient considered in this work is one order smaller than the one used in experiment~\cite{pol7}, thus the effective mass of the polariton is considered to be constant across the sample. Therefore the standard assumption about smallness of variation of the potential on one period due to gradient is valid, so that the influence of the gradient on the profiles of Bloch modes is negligible. The main effect of this potential is to trigger the variation of Bloch momentum of the wavepacket in the Brillouin zone, as discussed below. Since the very fact of existence of the edge states in topological polariton insulator is not connected with the presence of losses that are intrinsic in these systems and since Bloch oscillations is essentially linear physical phenomenon, we do not take losses into account in the model (1) and mention that such losses can be compensated by the external pump, see e.g. \cite{loss1,loss2,loss3,loss4}. Moreover, recent progress in technology of fabrication of high-Q microcavities with low losses \cite{loss5,loss6} enabled demonstration of long-living polariton condensates with lifetimes of several hundreds of picoseconds. Note that, experimentally,  the pump (either coherent or incoherent) is  switched off after the relatively low polariton and reservoir exciton densities are created so that any additional resonance shifts can be disregarded. Note that the spatial distributions in subsequent dissipative dynamics will not be affected by losses, except for overall decrease of density with time(I am not sure about this).

\section {\uppercase\expandafter{\romannumeral3}.Results and discussion}
The dynamics of Bloch oscillations is known to be strongly affected by the specific features of the Floquet-Bloch spectrum of the eigenmodes of the periodic structure at $\alpha=0$. Here we aim to elucidate the new phenomena introduced by the unusual spectrum of topological system. We  first set $\alpha=0$ in Eq. (1) and search for Bloch eigenmodes of the polariton topological insulator in the form $\boldsymbol{\psi}(x,y)=e^{iky-i\varepsilon(k) t}\boldsymbol{\phi} (x,y)$, where \(\varepsilon(k)\) is the energy, $k\in [0,\textrm{K}]$ is the Bloch momentum along the $y$-axis, $\textrm{K}=2\pi/Y$ is the width of the Brillouin zone, $\boldsymbol{\phi}(x,y)=\boldsymbol{\phi}(x,y+Y)$ is the periodic spinor function localized along the $x$-axis. The lowest part of the spectrum of our structure is shown in Fig.~1(e), for the case when simultaneous action of spin-orbit coupling (here we took $\beta=0.3$) and Zeeman splitting $\Omega=0.8$ results in the breakup of time-reversal symmetry in Eq. (1) and opening of the topological gap between the first and second spectral bands, which in the absence of the above mentioned physical effects would meet at two Dirac points at $k=\textrm{K}/3$ and $k=2\textrm{K}/3$. We deliberately selected a sufficiently large value of the Zeeman splitting to ensure a considerable separation in energy between the two depicted bands and the rest of the spectrum. This allows to considerably suppress Landau-Zener tunneling into higher bands. Due to the truncation of the topological insulator, two unidirectional in-gap edge states connecting two bands arise at $\textrm{K}/3<k<2\textrm{K}/3$ [green and red curves in Fig.~1(e)]. Edge states belonging to different branches are highly confined near the zigzag edges, when their energies $\varepsilon$ fall close to the center of the topological gap. See examples in Figs.~1(c),(d) corresponding to the points \textbf{c}, \textbf{d} in Fig.~1(e). However, they notably expand into the bulk of the array when the energy approaches the edge of the topological gap. For $k\to 0$ or $k\to\textrm{K}$ such modes smoothly transform into bulk states. Thus, the bulk state from Fig.~1(b) resides in the same continuous branch of the dispersion relation [point \textbf{b}] as the edge state from Fig.~1(c) [point \textbf{c}].

\begin{figure}[!hbt]
	\centering
	\includegraphics[width=\columnwidth]{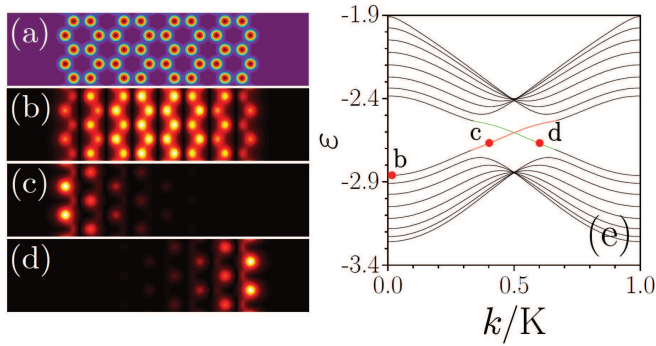}
	\caption{(a) Schematic illustration showing lattice of microcavity pillars with zigzag-zigzag edges and (b)-(d) examples of modulus distributions of dominating component $|\psi _{-}|$ in modes, corresponding to the points b-d on the energy-momentum diagram (e). In all cases $\beta=0.3$ and $\Omega=+0.8$.}
	\label{fig:one}
\end{figure}

In the presence of a potential gradient in Eq.~(1) the wavepacket experiences a constant force along $y$. If the force is small $(\alpha \ll 1)$, the evolution of the system is adiabatic: one can operate with the same set of eigenmodes, but under the action of the force the Bloch momentum of the wave in our narrow lattice ribbon slowly varies in time, $k(t)=k_0+\alpha t$, scanning the whole Brillouin zone \cite{bloch}. Therefore, the Bloch wave with a broad envelope and momentum $k_0$ moves along the corresponding branch of the dispersion relation, undergoing shape transformations in real space that reflect the modification of the wavepacket position in the spectrum from Fig.~1(e). Since the dependence $\varepsilon(k)$ is periodic, the evolution in the spatial domain is periodic, too, if the Landau-Zener tunneling to higher bands is weak \cite{zener}, which is the case in our system. If one uses for construction of broad wavepacket one of the modes from the depth of the first or second bands in Fig.~1(e), the wavepacket moves along the corresponding branch of the dispersion relation, remaining always in the bulk of the array, undergoing conventional BOs with period $T=\textrm{K}/\alpha$. The same standard dynamics of BOs (without any interband transitions and switching between different edges) is observed in nontopological system, where either spin-orbit coupling is set to zero $(\beta=0)$ or Zeeman splitting is absent $(\Omega=0)$ (recall that in such a system edge states are degenerate and there is no topological gap, hence wavepacket exciting mode from certain branch always remains in the same band).

\begin{figure*}[!htb]
\centering
{\includegraphics[width=\linewidth]{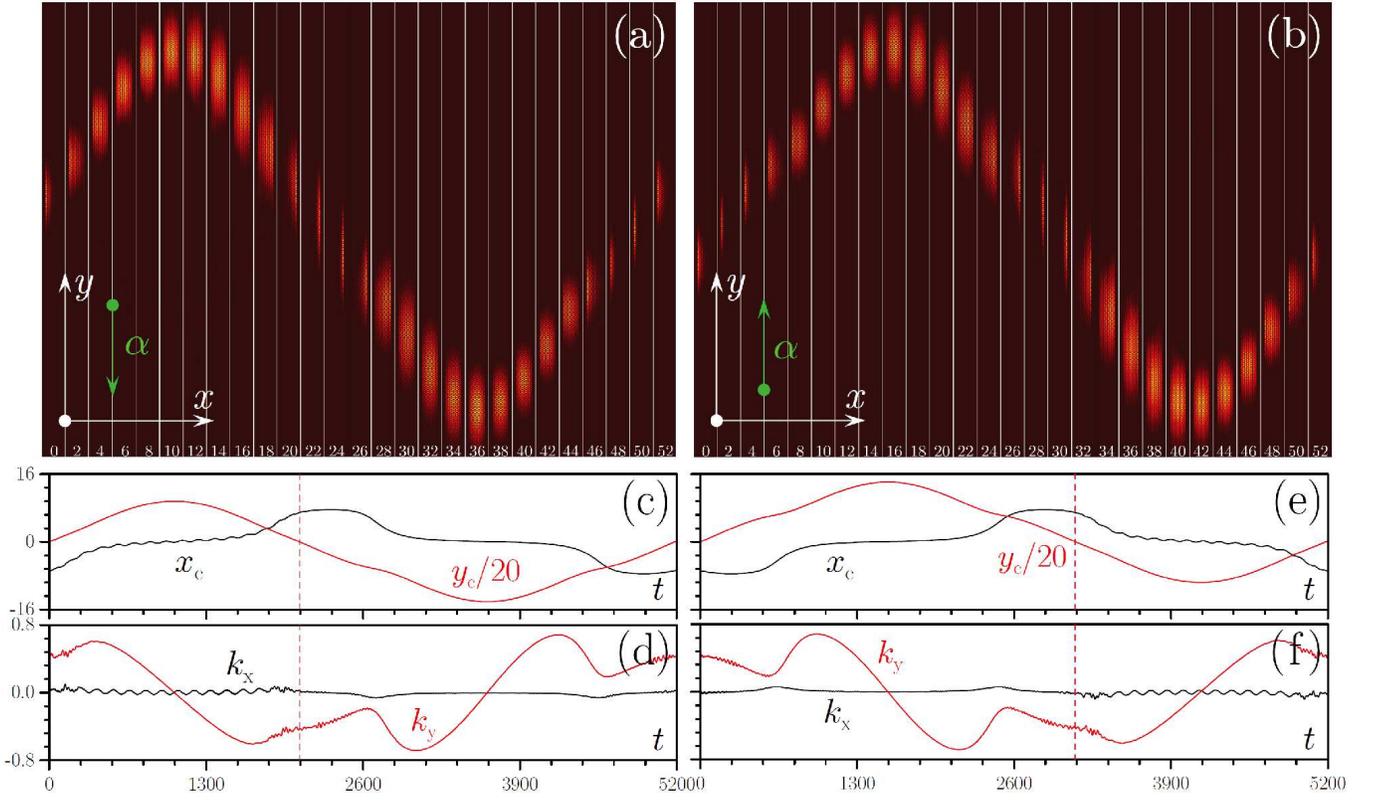}}
\caption{Distributions of $|\psi _{-}|$ in different moments of time corresponding to the number of the cross-section multiplied by $100$ showing dynamics of Bloch oscillations for positive Zeeman splitting $\Omega=+0.8$ and gradients $\alpha=-0.001$ (a), $\alpha=+0.001$ (b). Initially topological edge state with momentum $k=0.4\textrm{K}$ and width $w=30$ is located on the left edge. Green arrows indicate gradient direction. Red dashed lines in (c)-(f) indicate the moment of time when the wavepacket returns to the $y=0$ point, but at the opposite edge of the insulator.}
\label{fig:two}
\end{figure*}

The picture changes qualitatively when the wavepacket is constructed using topologically protected edge states at $\beta,\Omega \ne 0$ with broad $y$ envelope (of width $w=30$), such as the state with Bloch momentum $k_{0}=0.4 \textrm{K}$ corresponding to the point \textbf{c} in Fig.~1(e). Examples of the evolution dynamics are presented in Fig.~2. The selected edge state has positive group velocity $v(k)=d\varepsilon/dk$, and it moves in the positive direction of the $y$-axis. The state connects two different bands. Therefore, upon motion along the excited branch of the dispersion relation under the action of the constant force, the wavepacket traverses the topological gap and for $\alpha>0$ it transforms into a bulk state from the bottom of the second band. In real space this is accompanied by a considerable displacement along the $y$-axis and by a shift of the wavepacket into the bulk [see Fig.~2(b) at one quarter of the BO period]. Moving along the dispersion branch on the bottom of the second band, the wavepacket reaches the point $k=\textrm{K}$ and due to the periodicity of dispersion, reappears at $k=0$. In this point the group velocity changes its sign. Further variation of Bloch momentum induced by the force shifts the wavepacket back into the topological gap so that it reaches the point \textbf{d} corresponding to the edge state with negative group velocity and residing on the different edge [see Fig.~2(b) at half of the BO period].

Therefore, in the topological insulator, highly unconventional Bloch oscillations, involving switching between its opposite edges and periodic penetration into the bulk, occur. Despite the fact that the gradient is applied along the $y$-axis only, the wavepacket exhibits oscillations also along the $x$-axis, a remarkable phenomenon that is not known to occur in non-topological systems. After the point \textbf{d} is passed, the wavepacket transforms into the state from the top of the first band, i.e., it expands into the bulk again [see Fig.~2(b) at three quarters of the BO period]. After reaching the $k=\textrm{K}$ point, the wavepacket arrives to the point \textbf{b} and then comes back to the initial location \textbf{c} within the topological gap (i.e., it returns to the left edge in real space), completing one BO cycle. The described dynamics clearly shows that, in complete contrast to non-topological systems or to excitations in the depth of the band of topological system, BOs involving edge states exhibit a period $T=2\textrm{K}/\alpha$ that is two times larger than the period of usual BO. Thus, to return to the initial location the wavepacket has to traverse Brillouin zone twice.

\begin{figure*}[!htb]
\centering
\includegraphics[width=\linewidth]{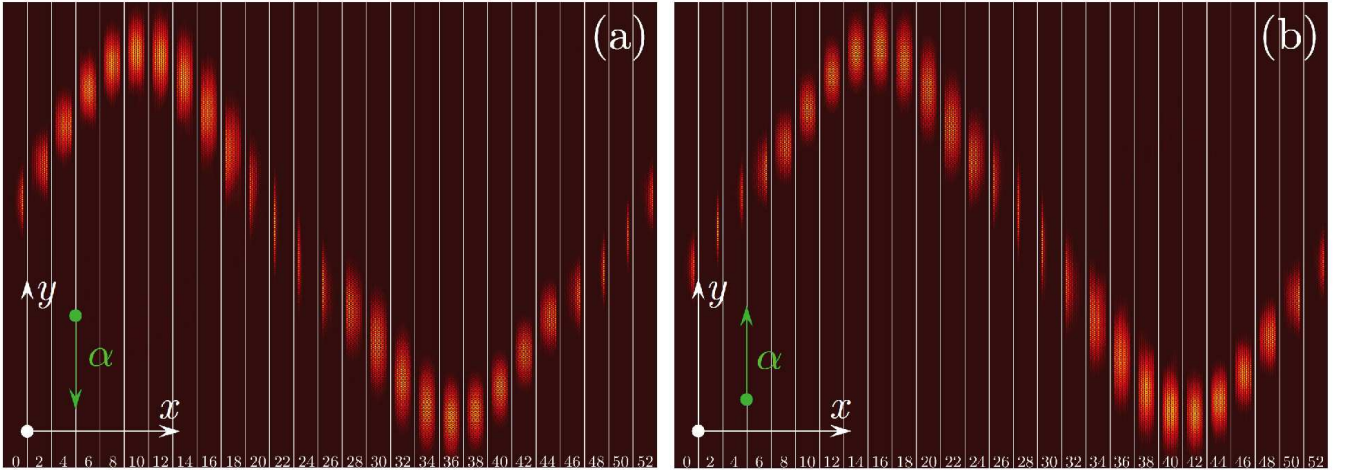}
\caption{Same as in Fig.~\ref{fig:two}, but for negative Zeeman splitting $\Omega=-0.8$ . Initially topological edge state with momentum $k =0.4\textrm{K}$ and width $w =30$ is located on the right edge. Evolution of wavepacket center and central momentum in the Fourier domain (not shown here) is identical to that shown in Fig.~\ref{fig:two}(c)-(f), but one has to change $x_{c}\rightarrow -x_{c}$, and $k_{x}\rightarrow -k_{x}$.}
\label{fig:three}
\end{figure*}

Note that if the input state corresponds to the edge state, an inversion of the sign of the gradient does not change the direction of BOs in real space (compare Figs.~2(a) and 2(b)). The direction of motion in the momentum space does change. Because for $\alpha<0$ the wavepacket upon evolution turns into the mode from the top of the first band (rather than into the mode from the bottom of the second band as it happened for $\alpha>0$) one can see that the structure of the wave in the bulk on the first and second halves of BO cycles is different for opposite gradients.

In Figs.~2(c)-2(f) we plot the coordinates of the center of mass of the wavepacket in real space $(x_c, y_c)$ and in the Fourier domain $(k_x, k_y)$ as functions of time, calculated by the expressions:

\begin{equation}
\begin{split}
(x_{c},y_{c}) = U^{-1}\iint (x,y)(|\psi_{+}|^2+|\psi_{-}|^2)dxdy,\\
(k_x,k_y) = 4\pi F^{-1} \iint (\kappa_x,\kappa_y)(|\tilde{\psi_{+}}|^2+|\tilde{\psi_{-}}|^2)d\kappa_xd\kappa_y,
\end{split}
\label{eq:cent}
\end{equation}

where $U=\iint (|\psi_{+}|^2+|\psi_{-}|^2) dxdy$, $F=\iint (|\tilde{\psi_{+}}|^2+|\tilde{\psi_{-}}|^2) d\kappa_xd\kappa_y$, and $\tilde{\psi_{+}}, \tilde{\psi_{-}}$ are the Fourier transforms of $\psi_{+}, \psi_{-}$. The oscillations of the $x$-coordinate of the wavepacket center are out-of-phase with oscillations of its $y$-coordinate, the amplitude of the latter  being much larger [Figs.~2(c) and 2(e)]. Notice that we study a relatively narrow topological insulator, to reduce the temporal period of the BO. The period can be drastically reduced by larger gradients $\alpha$, but this may lead to Landau-Zener tunneling. The periodic motion of the wavepacket in the spectral domain is readily visible in Figs.~2(d) and 2(f), which show a much larger variation in the $k_y$ component. The dependencies $k_{x,y}(t)$ are perfectly periodic, indicating that the wavepacket is almost exactly recovered after a BO cycle. Note, that the integral criterion \eqref{eq:cent} yields smooth time dependencies of $k_{x,y}$, see Figs. 2(d),(f), even when a wavepacket reappears at the other edge of the Brillouin zone.

A change of sign of the Zeeman splitting, from $\Omega=+0.8$ to $\Omega=-0.8$, significantly affects the dynamics of the BOs. The spectrum $\varepsilon(k)$ remains similar to that shown in Fig.~1(e), but with several noteworthy differences. First, inverting the sign of $\Omega$ changes the relative strength of the $\psi_{+}$ and $\psi_{-}$ spinor components. Second, the edge mode from the red (green) curve that resides at the left (right) edge of the array at $\Omega=0.8$, for the opposite sing of $\Omega$ it resides on the right (left) edge. Thus, if the mode at point \textbf{c} from Fig.~1(e) is excited, one starts the BO cycle from the mode on the right edge (see Fig.~3). The $y$-dynamics in this case remains the same, but evolution along the $x$-axis reverses: the trajectories of motion can be obtained from those shown in Fig.~2 if one changes $x_{c}\rightarrow -x_{c}$, and $k_{x}\rightarrow -k_{x}$. Thus, the conclusion is that the magnetic field that determines the direction of edge currents can also be used to change the $x$ component of the currents.

Figures 4(a) and 4(b) show the dependence of the complete time period $T$ and $y$-amplitude $A_y$ of BOs on the potential gradient $\alpha$ for a fixed spin-orbit coupling strength $\beta=0.3$. Here $T$ is defined as the time required for the wavepacket to return to the initial position after traversing twice the Brillouin zone, while $A_y$ is determined as a difference between the maximal and minimal $y$-positions of the wavepacket during evolution. In accordance with the model of adiabatic motion of the wavepacket within the Brillouin zone caused by a constant force described above, both these parameters vary as $\sim 1/\alpha$. While the period $T$ is independent of the spin-orbit coupling strength $\beta$, the amplitude of oscillations $A_y$ monotonically decreases with increasing $\beta$ [Fig.~4(c)]. Such a phenomenon was not expected. Indeed, the amplitude of BO is usually proportional to the maximal energy difference acquired by the wavepacket upon motion across the Brillouin zone and one may expect that the difference should grow with increasing $\beta$ due to broadening of the topological gap. However, while the gap broadens with $\beta$, the two lowest bands shrink, leading to an overall decrease of the interval of energies scanned by the wavepacket, which, in turn, leads to diminishing $A_y$.

\begin{figure}[!hbt]
		\centering
	\includegraphics[width=\columnwidth]{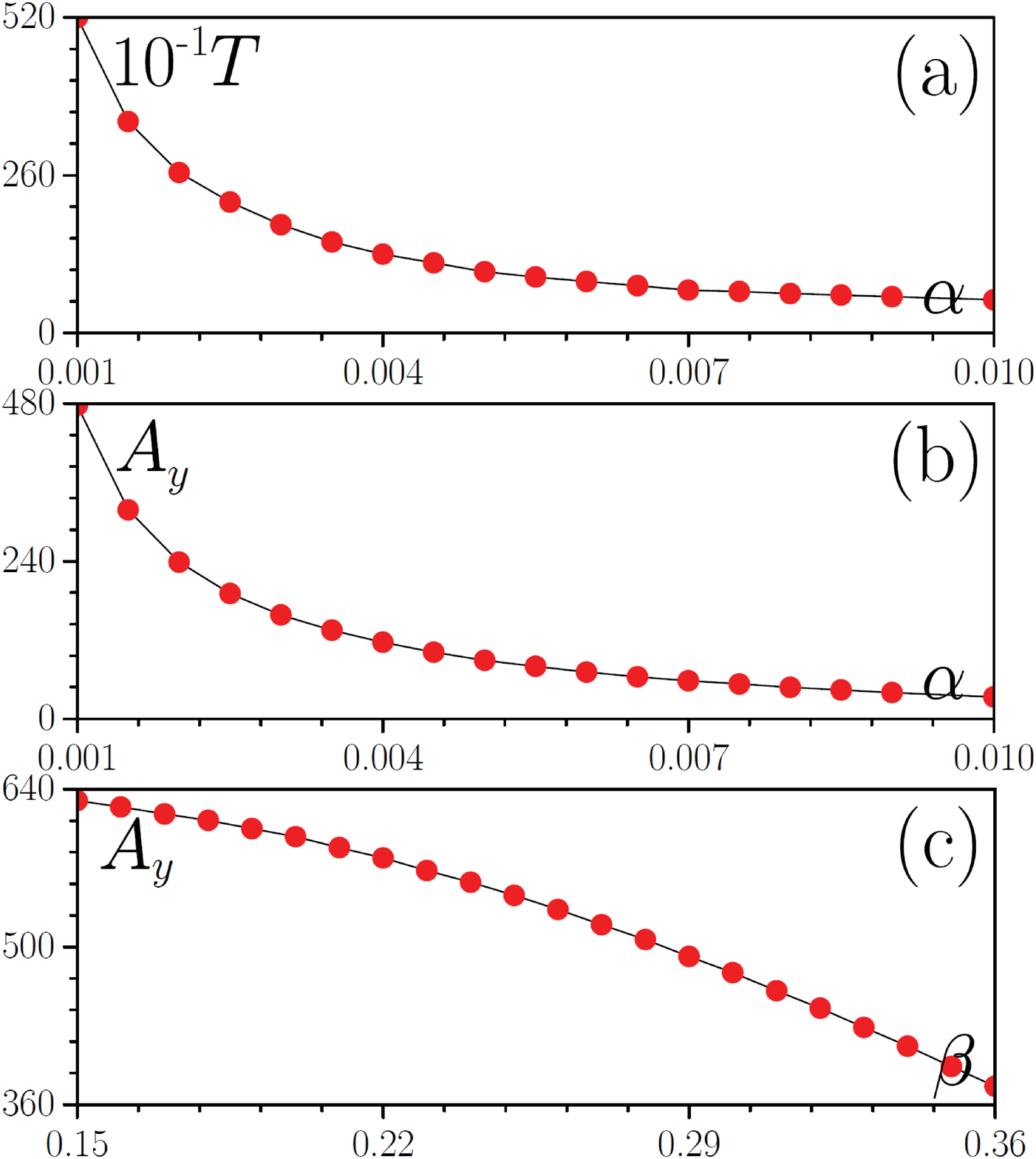}
	\caption{Period (a) and $y$-amplitude (b) of topological Bloch oscillations versus gradient $\alpha$ at $\beta=0.3$. (c) $y$-amplitude of Bloch oscillations versus strength of spin-orbit coupling $\beta$ at $\alpha=0.001$. In all cases $\Omega=+0.8$.}
	\label{fig:four}
\end{figure}

\begin{figure}[!hbt]
\centering
\includegraphics[width=\columnwidth]{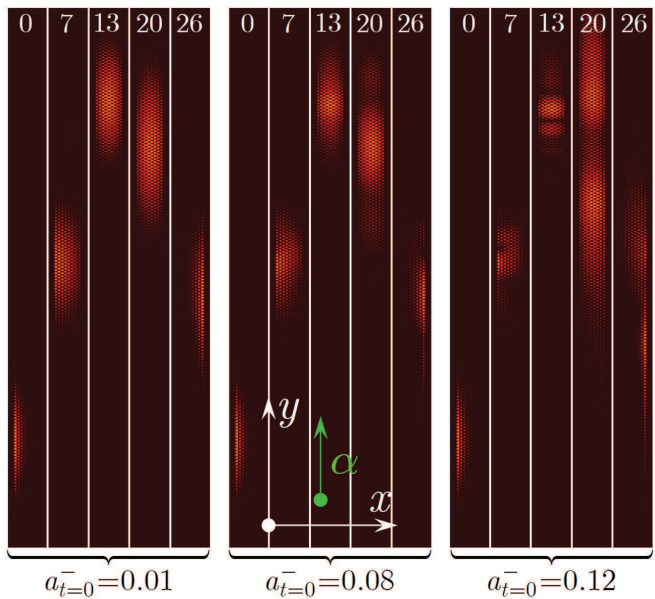}
\caption{Distributions of $|\psi _{-}|$ in different moments of time corresponding to the number of the cross-section multiplied by $100$ showing dynamics of \textit{nonlinear} Bloch oscillations for $\Omega=+0.8$, $\alpha=+0.001$, and different input peak amplitudes of the $\psi_-$ component indicated below panels. Half of Bloch oscillations cycle is shown. Initially topological edge state with momentum $k=0.4\textrm{K}$ and width $w=30$ is located on the left edge.}
\label{fig:five}
\end{figure}

Finally, we would like to stress that Bloch oscillations reported here can be observed even in the presence of nonlinear interactions in the low-density regime. To illustrate this we included corresponding nonlinear terms $(|\psi_\pm|^2+\sigma|\psi_\mp|^2)\psi_\pm$ accounting for repulsion between polaritons with the same spin and weak attraction $\sigma=-0.05$ between polaritons with opposite spins into right-hand side of the evolution Eq. (1). The dynamics of evolution within half of Bloch oscillations cycle for different input peak amplitudes $a^-_{t=0}$ of the dominating $\psi_-$ component in this nonlinear case is shown in Fig.~\ref{fig:five} for the same $\alpha,\Omega$ parameters as in Fig.~\ref{fig:two}(b). Bloch oscillations clearly persist up to amplitude values $a^-_{t=0}\sim 0.1$. For larger input amplitudes (hence stronger nonlinear effects) one observes distortions of the wavepacket and its splitting into two fragments. At the same time, the wavepacket still moves to the opposite edge of the ribbon after completing half of the oscillation cycle.

\section{\uppercase\expandafter{\romannumeral4}. Conclusion}

We presented a new type of Bloch oscillations, namely Bloch oscillations of topological edge states. The fundamental result that we have uncovered is that a full cycle of Bloch oscillation for a topological edge state is achieved through a continuous transformation between the localized edge mode and the delocalized bulk mode, as well as through a transition between the two edge states. In topological insulator the wavepacket traverses Brillouin zone twice to complete one Bloch oscillation cycle, thus the period of oscillations in topological system is two times larger than in the usual insulator. These topology-controlled phenomena are in sharp contrast to the behavior exhibited by non-topological systems.

\section{acknowledgments}
Y.V.K. and L.T. acknowledge support from the Severo Ochoa Excellence Programme (SEV-2015-0522), Fundacio Privada Cellex, Fundacio Privada Mir-Puig, and CERCA
/Generalitat de Catalunya. C.L. acknowledge support of the National Natural Science Foundation of China (NSFC) (Grant No. 11805145). Y.V.K. acknowledges funding of this work by RFBR and DFG according to the research project NO. 18-502-12080. F.Y. acknowledges support from the NSFC (Grants No. 61475101 and 11690033).

\bibliography{sciadvbib}

\bibliographystyle{ScienceAdvances}

\clearpage
\end{document}